\documentclass[journal=jacsat,manuscript=article, layout=twocolumn]{achemso}
\setkeys{acs}{keywords = true}
\setkeys{acs}{maxauthors = 0} 
\setkeys{acs}{articletitle = true}

\usepackage{chemformula} 
\usepackage[T1]{fontenc} 
\usepackage{cuted}

\author{Bing Cheng}

\affiliation[a]
{Department of Physics and Astronomy, The Johns Hopkins University, Baltimore, Maryland 21218, USA}

\author{T. Schumann}

\affiliation[b]
{Materials Department, University of California, Santa Barbara, California 93106-5050, USA}

\author{Youcheng Wang}

\affiliation[a]
{Department of Physics and Astronomy, The Johns Hopkins University, Baltimore, Maryland 21218, USA}

\author{X. Zhang}

\affiliation[a]
{Department of Physics and Astronomy, The Johns Hopkins University, Baltimore, Maryland 21218, USA}

\author{D. Barbalas}

\affiliation[a]
{Department of Physics and Astronomy, The Johns Hopkins University, Baltimore, Maryland 21218, USA}

\author{S. Stemmer}

\affiliation[b]
{Materials Department, University of California, Santa Barbara, California 93106-5050, USA}

\author{N. P. Armitage}
\email{npa@jhu.edu}

\affiliation[a]
{Department of Physics and Astronomy, The Johns Hopkins University, Baltimore, Maryland 21218, USA}

\title {A Large Effective Phonon Magnetic Moment in a Dirac Semimetal}

\keywords{Topological semimetal, Cd$_3$As$_2$ thin film, magneto-THz, Dirac fermion, phonon magnetic moment}

\begin{document}

\begin{abstract}
 We investigated the magnetoterahertz response of the Dirac semimetal  Cd$_3$As$_2$ and observed a particularly low frequency optical phonon, as well as a very prominent and field sensitive cyclotron resonance.  As the cyclotron frequency is tuned with field to pass through the phonon, the phonon become circularly polarized as shown by a notable splitting in their response to right- and left-hand polarized light. This splitting can be expressed as an effective phonon magnetic moment that is approximately 2.7 times the Bohr magneton, which is almost four orders of magnitude larger than ab initio calculations predict for phonon magnetic moments in nonmagnetic insulators.  This exceedingly large value is due to the coupling of the phonons to the cyclotron motion and  is controlled directly by the electron-phonon coupling constant.  This field tunable circular-polarization selective coupling provides new functionality for nonlinear optics to create light-induced topological phases in Dirac semimetals.
\end{abstract}


\section{Introduction}

A number of linear and nonlinear magneto-optical effects from relativistic fermions and Berry curvatures are anticipated in 3D topological semimetals (TSMs) \cite{NPA18}.  Besides their appealing electronic features, the interplay between electronic states and other degrees of freedom, such as lattice vibrations and magnons, have also begun to attract attention \cite{Phonon_chiral_2016,Phonon_chiral_2017,Circular_phonon_Dichroism_2017,Mn3Sn_photoemission_2017}.  For instance, it has been predicted that in a Weyl semimetal, the "chiral current" which corresponds to the transfer of charge between Weyl nodes, can interact with $A_1$ Raman-active phonon modes and make them infrared-active \cite{Phonon_chiral_2016,Phonon_chiral_2017}. Such phonon modes could also hybridize with the plasmon mode of Weyl fermions to form an anticrossing structure in a similar fashion to the Kondo hybridization in heavy-fermion systems \cite{Magneto_phonon_1991,GaAs_quantum_well_phonon_1997}. 

In this work, we report the observation of another interesting charge-phonon effect in epitaxial thin films of the Dirac semimetal Cd$_3$As$_2$.  The temperature-dependent terahertz (THz) conductivity exhibits coherent metallic transport and a sharp optical phonon mode.  Due to the low cyclotron mass ($\sim$ 0.03$m_e$) in this Dirac system, a sharp cyclotron resonance mode develops and moves rapidly to higher frequency with increasing field.  As the cyclotron resonance mode is tuned to pass through the optical phonon frequency, the phonon becomes circularly polarized with a large splitting in the energies of right- (\textit{r}-) and left- (\textit{l}-hand) polarized phonons, accompanied by a notable Fano asymmetry of the \textit{l}-hand branch.  The splitting is almost four orders of magnitude larger than the prediction from first-principle calculations of the phonon "Zeeman effect" in undoped semimetals\cite{juraschek2018orbital} and leads to an effective phonon magnetic moment of almost 2.7 Bohr magnetons -- an unprecedentedly large value in a nonmagnetic system.  We attribute this large enhancement to the interaction between the circularly polarized cyclotron motion and the lattice degrees of freedom. In this regard the size of the moment is set directly by the electron-phonon coupling constant and -- as we show -- provides a measure of it.  This field controllable and circular polarization selective coupling also provides new opportunity for nonlinear optical methods to induce and study exotic light-induced topological phases.

\section{Results and discussion}

\textbf{Sample details.} Cd$_3$As$_2$ is a prototype of 3D Dirac semimetal with $I4_1/acd$ lattice structure which has been the subject of extensive recent studies\cite{ali2014crystal}.  It has a pair of four-fold degenerate Dirac nodes located along the $k_z$ axis. The Dirac nodes are protected by a C$_4$ symmetry around the $z$ axis and cannot be removed except by breaking this symmetry.  Recently, high quality epitaxial Cd$_3$As$_2$ films grown by molecular-beam epitaxy have become available\cite{Timo_CdAs_growth_16,CdAs_MBE_2}.  The films in this work were grown on (111)B GaAs substrates to a thickness of 280 nm with the Cd$_3$As$_2$ 112 direction normal to the surface. Magnetoterahertz measurement was performed by time-domain magneto-THz spectroscopy (Fig 1a).  Further details of these measurements and analysis are provided in Ref. \cite{TCI_Bing_PRL} and the \textit{Supporting Information (SI) Note 1, 2, and 3} .

\begin{figure*}[t]
\centering
\includegraphics[width=1\linewidth]{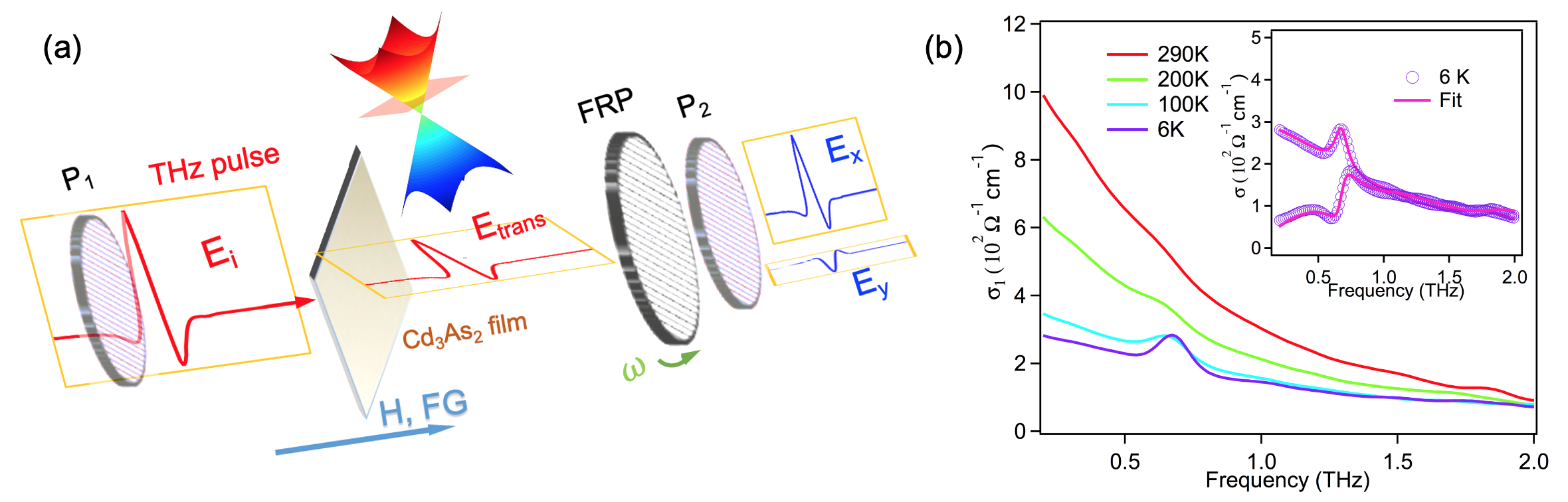}
\caption{ (a) Schematic of the time-domain magnetoterahertz spectrometer. A wire grid polarizer P$_1$ is placed in front of the sample to produce a vertically polarized terahertz pulse $E_i$. The yellow frames represents the polarization planes of the terahertz pulses. Upon transmission, the polarization may be rotated as shown by the tilted yellow frame.  The fast rotating polarizer method (FRP) is used to modulate the transmitted terahertz pulse at a frequency $\omega/2 \pi$ = 47 Hz. With a second wire grid polarizer P$_2$ behind the FRP, the transmitted terahertz pulse can be decomposed into two orthogonal linearly polarized terahertz pulses along vertical ($E_{x}$) and horizontal directions ($E_{y}$) by using a lock-in amplifier locked-in at a frequency $2 \omega$. By performing similar measurements on a reference, the complex transmission matrix elements $T_{xx}$ and $T_{xy}$ can be determined through a single measurement to high precision (see more details in \textit{SI Note 2 and 3}) and Ref. \cite{Morris12a}.  (b) Real parts of THz conductivity of Cd$_{3}$As$_{2}$ film at four temperatures. Inset shows the Drude-Lorentz fit to the real and imaginary parts of THz conductivity at 6 K.}

\end{figure*}

\textbf{Terahertz Conductivity at Zero Field.} The real part of the zero-field THz conductivity $\sigma_{1}$ is displayed in Fig 1b. At 6 K, $\sigma_{1}$ shows a Drude-like peak with a sharp phonon mode at 0.67 THz.  Our analysis in \textit{SI Note 5} shows this to be a doubly degenerate $E_u$ zone center optical phonon.  With increasing temperature, the Drude part of $\sigma_{1}$ becomes larger and sharper and the phonon mode becomes broader. A Drude-Lorentz fit for real and imaginary parts of conductivity at 6 K are shown in the inset of Fig 1b (See the details of fitting formula in \textit{SI Note 4}).  This fit includes s Drude term and a finite frequency Lorentz term. One can see the THz conductivity at 6 K is well reproduced by this fitting with a Drude plasma frequency ($\omega_p$/2$\pi$) and scattering rate (1/2$\pi\tau$) $\sim$22 THz and $\sim$0.9 THz, and the plasma frequency and linewidth of the phonon $\sim$4.5 THz and $\sim$0.12 THz. The weak broad features above 1.6 THz probably arises from other higher energy phonon modes with large damping.  By careful inspection of the lineshape of the phonon at 0.67 THz, we can see that the phonon mode exhibits a weak asymmetry which indicates it has a detectable coupling to the continuum of electronic states. This will be discussed in more detail below.

\begin{figure*}[t]
\centering
\includegraphics[clip,width=6.3in]{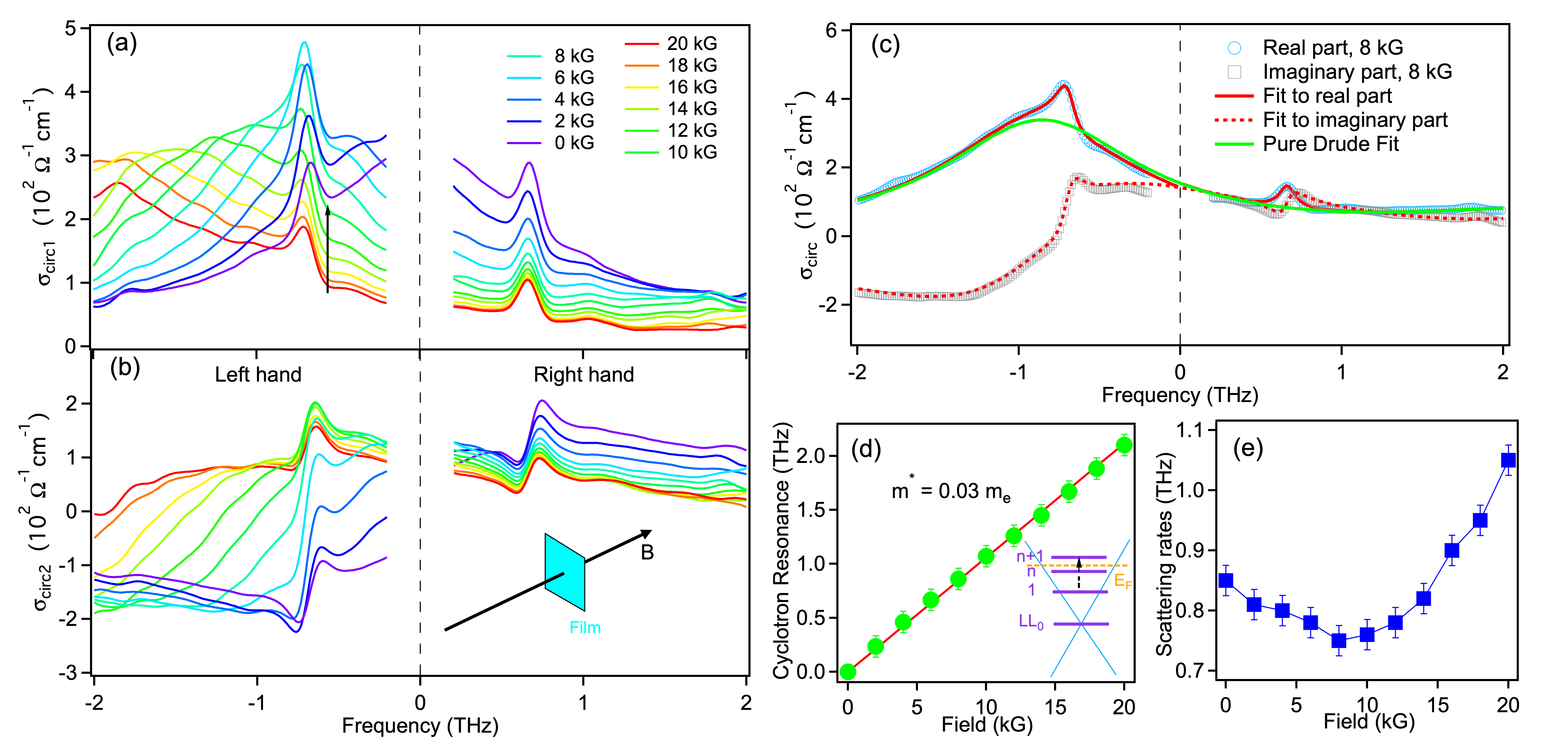}
\caption{(a) Real and (b) Imaginary parts of the magneto-THz conductivity under circular basis at 6 K. \textit{r}-hand and \textit{l}-hand THz conductivity are displayed as positive and negative frequencies respectively. The inset shows the configuration of the sample and magnetic field in Faraday geometry (c) Drude-Lorentz fits of real and imaginary parts of THz conductivity under circular basis at 8 kG. (d) Cyclotron resonances as a function of field. Inset shows intraband inter-LL transitions when the Fermi level is located between LL$_{n}$ and LL$_{n+1}$. (e) Drude scattering rates as a function of field at 6 K. }
\label{xxx}
\end{figure*}

\textbf{Circularly Polarized Terahertz Conductivity.} The THz conductivity in the circular polarization basis was measured by using the fast rotating polarizer method \cite{TCI_Bing_PRL}.  See \textit{SI Note 3} for technical details. In Fig 2a and 2b, we show the real and imaginary parts of the THz conductivity in the circular polarization basis ($\sigma_{circ}$) at different fields.  It is quite illustrative to display the response to the \textit{r}- and \textit{l}-hand polarized light as positive and negative frequencies respectively. This follows from the fact that we may understand \textit{r}- and \textit{l}-hand polarized light as having time dependencies that go as $e^{ \mp i \omega t}$. Hence the conductivity becomes a single continuous function of frequency that smoothly extends through zero frequency.  At zero field, the real part of $\sigma_{circ}$ is a function peaked at zero frequency, which is a typical metallic response. With increasing positive field, the peak moves quickly to finite negative frequency, while the conductivity is suppressed on the positive frequency side. This large shift of the peak with relatively small magnetic field can be identified as the cyclotron resonance (CR) mode of the $n$-type carriers [Inset of Fig 2d] with a very small cyclotron mass which arises because of the system's Dirac nature.  One of the most interesting aspects of $\sigma_{circ}$ is the field evolution of the $\pm$ 0.67 THz phonon. One can see that in the \textit{r}-hand channel the phonon's peak position decreases and lineshape changes with increasing field, while the \textit{l}-hand phonon shows a first similar movement but to higher frequencies and then at higher fields an even larger response to field. As shown in Fig 2a, the low frequency side of the \textit{l}-hand phonon develops a weak dip around $-$0.6 THz marked by the black arrowed line. This weak feature cannot be interpreted in terms of electronic excitations alone and comes from a "Fano" asymmetry induced by magnetic field-enhanced electron-phonon coupling. Fano resonance is a general phenomenon that arises from a interference between a sharp mode and a continuum background.

To separate electronic and phonon components, we used a conventional Drude/Drude-Lorentz model to fit the complex \textit{r}- and \textit{l}-hand THz conductivities simultaneously.  The total THz conductivity is $\sigma_{circ} = \sigma_{Drude} + \sigma_{phonon} $ plus a very weak electron-like oscillator that allows the spectrum to be fit on the positive frequency side at high fields. Its incorporation has no impact on the conclusions of the paper.  As magnetic field is applied, the zero frequency Drude response shifts to finite frequency and gives the distinct cyclotron resonance.  The expression for the phonon conductivity incorporates Fano effects.  Details of fitting formulas can be found in the \textit{SI Note 4}.

In Fig 2c, we show a Drude-Lorentz fit for the conductivity at 8 kG. One can see that the fit reproduces well the conductivity over the whole frequency region. We show the CR frequency as a function of field in Fig 2d. The CR linearly increases with field. Although Cd$_3$As$_2$ is a 3D Dirac semimetal and its carriers are massless Dirac fermions, under these weak fields the system's response is semi-classical and its CR gives the classical linear field evolution: $eB$/$m^{*}$, where $m^*$ is the cyclotron mass ($m^* = \hbar k_F / v_F$ without interactions and in a linear dispersion approximation).  By fitting the field dependence of the CR, $m^{*}$ is found to be 0.03 free electron masses, which is in agreement with the mass from temperature-dependent Shubnikov-de Haas oscillations \cite{goyal2018thickness}. The small value of $m^{*}$ arises from the low chemical potential of these TSMs.  Combining with the extracted zero field scattering rate, the electronic mobility can be estimated to be $10^4$ cm$^{2}$V$^{-1}$s$^{-1}$, consistent with previous dc transport measurement \cite{Dirac_Timo_prb}. In Fig 2e we plot the scattering rate of the Drude oscillator as a function of field. With magnetic field, the scattering rate initially decreases before increasing above 8 kG. This crossover is a unique feature that has not been reported in TSMs before. It is well known that the dc magneto resistivity of TSMs usually has a very complicated field dependence\cite{Cd3As2_magneto_resis_2014,magnetoresistivity_pnas,Dirac_Timo_prb}. Most discussion of its field dependent behavior assumes the scattering rate of Dirac/Weyl fermions is field independent. The crossover we observed in Cd$_3$As$_2$ provides new and important insights for the community to understand the complicated dc magneto transports of TSMs. Moreover, we will show below that this crossover comes from the competition between the decreasing impurity scattering\cite{TCI_Bing_PRL} (Dirac fermions and impurities) and the increasing field-tuned electron-phonon scattering (Dirac fermions and the phonon at 0.67 THz).

\begin{figure*}[t]
\centering
\includegraphics[clip,width=5.5in]{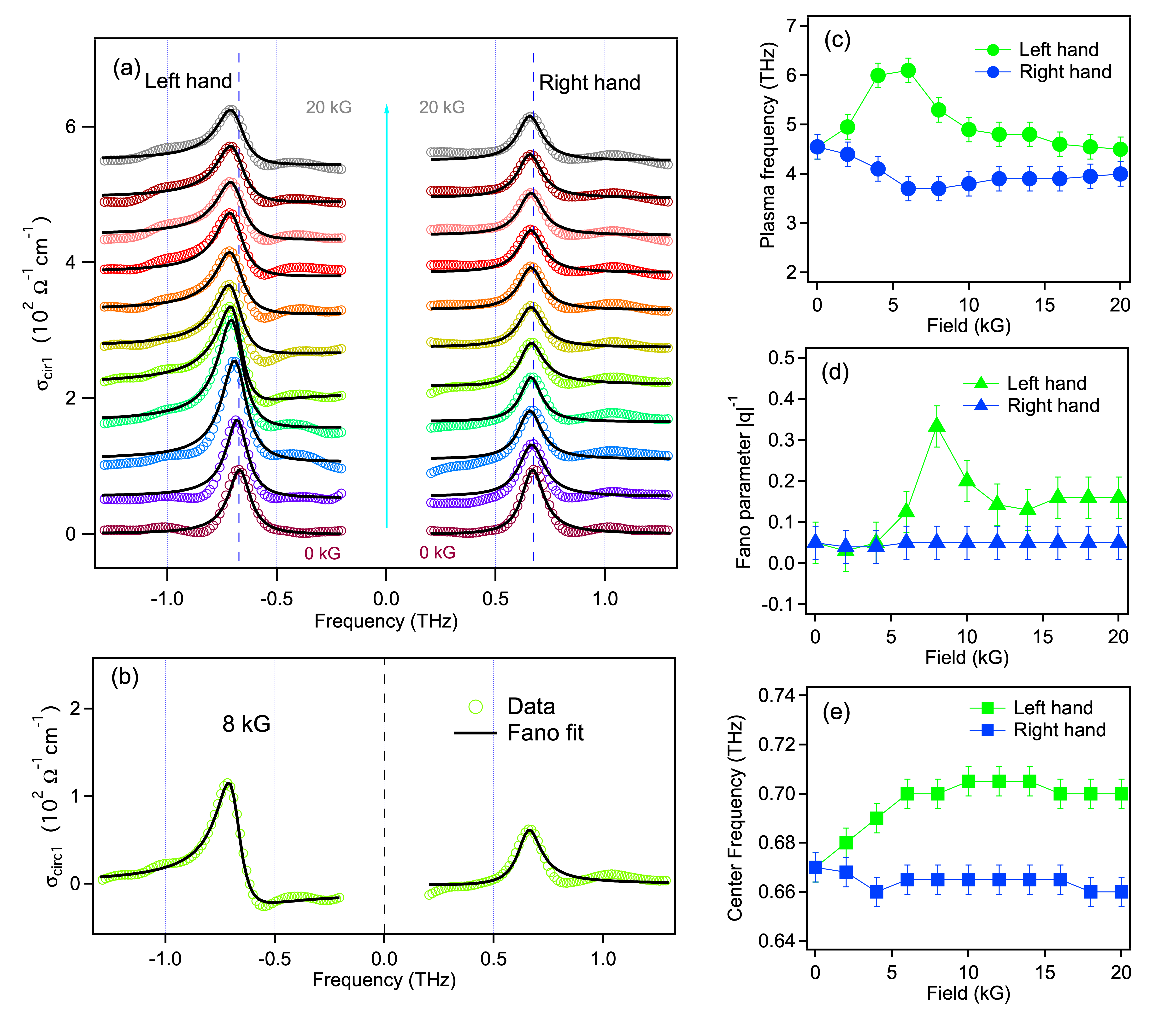}
\caption{(a) The real part of the THz conductivity of the phonon resonance after subtracting the electronic Drude background under circular basis at 6 K. \textit{r}- and \textit{l}-hand THz conductivity are displayed as positive and negative frequencies with the same offset between each. The colored makers are experimetal data and the black solid curves are the fits. (b) Fano fit for the phonon at 8 kG. (c) Plasma frequency $\Omega_p$ and (d) Fano parameter $\left|q\right|$$^{-1}$ and (e) Center frequency $\omega_0$ of the phonon as a function of field.  }
\label{Fig3}
\end{figure*}


\textbf{Circularly Polarized Phonon Mode.} To exhibit the field evolution of the phonons clearly, we subtract the pure electronic signal (the green curve in Fig 2c) and plot the THz conductivity of \textit{l}- and \textit{r}-hand phonons in Fig 3a with offsets. The markers represent the experimental data and the black curves are the Fano fits to the phonons. The \textit{l}-hand phonon develops a clear asymmetry with increasing field, while the \textit{r}-hand phonon shows smaller changes. To see this point more clearly, in Fig 3b, we show the Fano fit to the \textit{l}- and \textit{r}-hand phonons at 8 kG. The formula details of THz conductivity with Fano asymmetry can be found in \textit{SI}. We can see that the fit captures all features of both channels.  Besides the Fano asymmetry, the plasma frequency $\Omega_p$ of the \textit{l}-hand phonon becomes larger than $\Omega_p$ of the \textit{r}-channel. In Fig 3c and 3d, we show the field-dependence of $\Omega_p$ and the Fano parameter  $\left|q\right|$$^{-1}$ (that parametrizes the asymmetry in the phonon lineshape) for \textit{l} and \textit{r}-hand phonons. One can see that both $\Omega_p$ and $\left|q\right|$$^{-1}$ of the \textit{l}-hand phonon shows a resonance enhancement around 6 $\sim$ 8 kG. In contrast, in the \textit{r}-channel an enhancement is not observed. The phonon frequencies $\omega_0$ show distinct field dependencies. As shown in Fig 3e, $\omega_0$ in the \textit{l}-channel increases quickly from 0 to 8 kG and then stays constant. In contrast, $\omega_0$ in the \textit{r}-channel shows a small initial decrease before saturating. Near 6 kG, the splitting of phonons $\Delta\omega/2 \pi$ is 0.04 THz. The resonance features presented in \textit{l}-hand phonon at $\sim$6 kG strongly indicate the CR mode plays an important role in the asymmetry between \textit{l}- and \textit{r}-hand phonon because the CR energy crosses the phonon's central frequency when the field is 6 - 8 kG.

\textbf{Electron-Phonon Coupling.} For this case of coupling of a phonon to an electronic continuum, the Fano parameter $q$ is determined by the expression $ q^{-1}=\pi D_{eh}g_{ep} \frac{\mu_{eh}}{\mu_{ph}}$~\cite{Bilayer-graphene_Fano_2009}.  Here D$_{eh}$ is the electronic joint density of states that arises from the electron-hole pair intraband inter-LL transitions near the phonon frequency $\omega_0$. $g_{ep}$ is the electron-phonon coupling strength, and $\mu_{ph}$ and $\mu_{eh}$ are the optical matrix elements of phonons and electron-hole pairs respectively.  The electron-phonon coupling strength $g_{ep}$ is not expected to have a strong field dependence, but obviously D$_{eh}$ will.   As shown in Fig 2a, optical conductivity on the negative frequency side is gradually enhanced but on the positive frequency side it is suppressed as the CR moves in the negative frequency direction. D$_{eh}(\omega_0)$ would reach its maximum when the CR resonates with the phonon in the \textit{l}-hand channel.  This is presumably why the phonon plasma frequency and the Fano parameter in the \textit{l}-hand channel show a resonance structure near $\omega_0$.  It is also straightforward to understand the mechanism for the increasing field dependence of Drude scattering rate [Fig 2e] as a field-enhanced coupling between the optical phonon and massless Dirac fermions. Above 8 kG, the magnetic field enhanced electron-phonon scattering surpasses the decreasing trend of LL broadening from impurity potentials\cite{TCI_Bing_PRL}, and modifies the scattering rate to be an increasing function of field.

\textbf{Large Phonon Magnetic Moment.} Our most important observation is the splitting of the phonon into $r$ and $l$ polarization branches. This may be the largest splitting of phonons ever recorded due to the effects of time-reversal symmetry breaking. Extensive experiments in the 1970s showed in insulating \textit{magnetic} crystals a splitting between $r$ and $l$ polarized phonon branches could be triggered by a magnetic field. For instance, Schaack showed in strongly paramagnetic CeF$_3$ a two-fold degenerate optical phonon at 49.5 meV can be split by as much as $\sim$0.5 meV at 7 kG \cite{schaack1976observation}, leaving the relative splitting $\Delta\omega$/$\omega_0$ $\sim$0.01. As far as we know, this is the largest phonon splitting ever measured in absolute energy scale, but note even in absolute energy unit this splitting is only twice as large that what we measured at 7 kG in Cd$_3$As$_2$. However, a few points must be kept in mind to understand why our finding should be considered exceptional. First, the phonon in CeF$_3$ is itself almost 20 times the energy of the one we are considering in Cd$_3$As$_2$, which pushes the overall scale of the effect to much higher energies. The relative splitting $\Delta\omega$/$\omega_0$ in our case is 0.06, much higher than the value reported in CeF$_3$.  Second, CeF$_3$ is a magnetic system and small applied field aligns spins to greatly enhance the effects of applied magnetic field.

\textbf{Discussion.} The large splitting between $r$- and $l$-hand modes can be understood as follows.  In uniaxial crystals, zone center phonons are doubly degenerate and polarized in the plane normal to the $z$-axis.  Quite generally, these modes become \textit{r}- and \textit{l}-hand polarized and split linearly with field when a magnetic field is applied with a component along the symmetry axis \cite{anastassakis1972morphic}.  For small fields one can show that $\omega_j  \approx \sqrt{K_{xx}^e} \pm  \frac{K_{xyz} B_z}{2 \sqrt{K_{xx}^e}}$, where  $K_{xx}^e $ and $ K_{xy} =  K_{xyz} B_z$ are effective spring constants for motion in the $x-y$ plane.   The eigenstates for motion with nonzero magnetic field are $(1/\sqrt{2} ) (1, \pm i ,0) $. See \textit{SI} for further details.  Because of their circular polarization and splitting of eigenfrequencies, an orbital magnetic moment related to $  K_{xyz} $ can be assigned to the phonons.  There has been related recent (and earlier) interest in systems where phonons can be imbued with characteristics found in other lattice excitations such as magnetic moments, angular momentum and Berry phase structures \cite{phonon_Berry_2011,zhang2015chiral}. Juraschek and Spaldin used density functional theory to study the field-induced phonon splitting in \textit{nonmagnetic} compounds and found the relative splitting  ($\Delta\omega/\omega_0$) in most nonmagnetic compounds would be $\sim$ $10^{-6}$ to $10^{-4}$ \cite{juraschek2018orbital}.  Zhang and Niu showed that in inversion symmetry broken 2D chalcogenides that valley phonons could possess intrinsic angular moment in the finite (but opposite) angular momentum in the two $K$ and $K'$ valleys \cite{zhang2015chiral}. This has recently been confirmed in optical pump-probe spectroscopy experiments \cite{zhu2018observation}.

As mentioned above, the linear field dependence of the Cd$_3$As$_2$ phonons at small field can be regarded as an effective orbital moment. The relative energy splitting $\Delta \omega/\omega_0$ in weak magnetic field is $\sim$0.06 and equivalent to a magnetic moment of 2.5 $\times 10^{-23}$ m$^2 \cdot $A at 6 kG, which is 2.7 Bohr magnetons.   This extremely large value is approximately 3 to 4 orders of magnitude larger than predicted in nonmagnetic insulators\cite{juraschek2018orbital} where the size of the effect is set by the ionic cyclotron mass $eB/M$ (where $M$ is an effective ionic mass).  Therefore it is reasonable to ascribe the large splitting of phonon frequencies in this case and their circular polarization to the resonant enhancement of the cyclotron motion circulating with or against the circular motion of the phonons.   A minimal model \cite{xiaoguang1985two,peeters1986cyclotron,goerbig2011electronic,goerbig2007filling} for coupling the cyclotron motion to Einstein phonons at $\pm \omega_0$ considered in the \textit{SI Note 5} shows that in the small field limit the cyclotron resonance gets modified as $E_{cr} = \hbar \omega_{cr} \frac{ 1 }{1 + \lambda} $ and the phonons split as $E_{r,l} = \pm \hbar \omega_{0} \sqrt{ 1 + \lambda/2} + \hbar \omega_{cr} \frac{ \lambda/2 }{ 1 + \lambda } $, where $\lambda = 2 g_{ep}^2 / (\hbar \omega_{0})^2 $ is the dimensionless electron-phonon coupling constant. Please note these equations from the perturbation treatment are only valid when $\omega_{cr}$ is smaller than $\omega_{0}$. For the cyclotron resonance, our model reproduces the known result that the cyclotron mass is renormalized by the factor $1 + \lambda$ (See \textit{SI Note 5}).  This minimal model demonstrates a novel mechanism for generation of large phonon magnetic moments through electron-phonon coupling.  And it gives the remarkable result that at small fields the difference between the phonon splittings divided by the cyclotron resonance frequency is precisely the electron-phonon coupling constant $\lambda$.   As applied to our data, we find a dimensionless electron-phonon coupling constant of approximately 0.09.

\section{Summary and Outlook}

Our work features the first terahertz observation for a cyclotron resonance mode, an optical phonon mode, their magnetic field-enhanced interaction, and a large effecive phonon magnetic moment simultaneously in the nonmagnetic Dirac semimetal Cd$_{3}$As$_{2}$. However, we believe our findings are not limited to this particular case; the general idea should be more widely applicable in other TSMs.  By tuning their charge densities via gating TSM film devices, one may also study the gating- and field-tunable charge-phonon coupling in these TSMs.  Aside from being interesting in their own right, the observation of these rich charge and lattice dynamics and their response to magnetic field in Cd$_3$As$_2$ may provide new pathways to study novel light-induced phases in TSMs by ultrafast manipulation of lattice degrees of freedom \cite{light_induced_topo_phase_2017}. Among other aspects, one may use a narrow-spectrum multicycle intense THz pump pulse resonating with the phonon to excite a Cd$_3$As$_2$ film. The strong stimulus of the phonon will break  symmetries that may drive the Dirac semimetal into a light-induced topological insulator phase. Moreover, one may use intense circularly polarized pump pulse to excite the sample and drive it into a light-induced Floquet-Weyl semimetal phase where the fourfold degenerate Dirac node is split into two separate Weyl nodes.

\begin{acknowledgement}

Experiments at JHU were supported by the Army Research Office Grant W911NF-15-1-0560.  Work at UCSB was supported by the Vannevar Bush Faculty Fellowship program by the U.S. Department of Defense (grant no. N00014-16-1-2814).   We would like to thank Z. Tagay for help with some of the matrix analysis that now appears in the \textit{SI}.

\end{acknowledgement}

\begin{suppinfo}

 Time-domain magnetoterahertz spectroscopy, raw time trace data of substrate and sample at 6 K, calculations of optical conductivity in circular polarization basis, the formulas to fit terahertz conductivity, circularly polarized eigenstates of phonon by applied magnetic field

\end{suppinfo}


\newpage


\section{S\lowercase{upporting} N\lowercase{ote} 1:  T\lowercase{ime-domain} TH\lowercase{z spectroscopy}}

\renewcommand{\thefigure}{S\arabic{figure}}

Complex values of the transmission matrix $T_{xx}$ and $T_{xy}$ in THz range were measured by a home-built time-domain THz spectrometer in a closed-cycle 7 T superconducting magnet by Fourier transforms of the time-domain signals discussed above.   GaAs Auston switches are used as emitters and receivers to generate and detect THz pulses. An ultrafast laser (800 nm) is split into two paths by a beamsplitter. One beam travels to the biased emitter and generates a THz pulse. This THz pulse passes through the sample or substrate, modified, and arrives at the receiver. The other laser beam propagates to the receiver and is used to gate the THz pulse after passing through the sample. The beam path difference between these two laser beams is precisely controlled by a delay stage to map out the E field as a function of time of the THz pulse. By mapping out THz pulses after transmitting through substrates and samples separately, and taking a ratio of the Fourier transforms, we obtain a transmission function in the frequency domain. The complex conductivity of the thin films can be directly extracted in the thin-film limit with the expression: $T(\omega) = \frac{1+n}{1+n+Z_{0}d\sigma(\omega)}\mathrm{exp}[\frac{i\omega}{c}(n-1)\Delta L]$. Here $T(\omega)$ is the transmission of a particular eigenpolarization as referenced to GaAs substrate, $\sigma$($\omega$) is the complex optical conductivity in the corresponding basis, $d$ is the film thickness, $n$ is the index of refraction of the substrate and $Z_{0}$ is the vacuum impedance.  As discussed in main text, a fast rotating polarizer (FRP) setup was used to modulate the polarization of THz pulses being transmitted through the sample, allowing the polarization of the pulse to be determined to high accuracy in a single measurement \cite{Morris12a}. With the knowledge of the polarization state of the transmitted THz pulse, one can calculate the optical conductivity in the circular basis as described below.

\begin{figure*}[t]
\includegraphics[clip,width=5.5in]{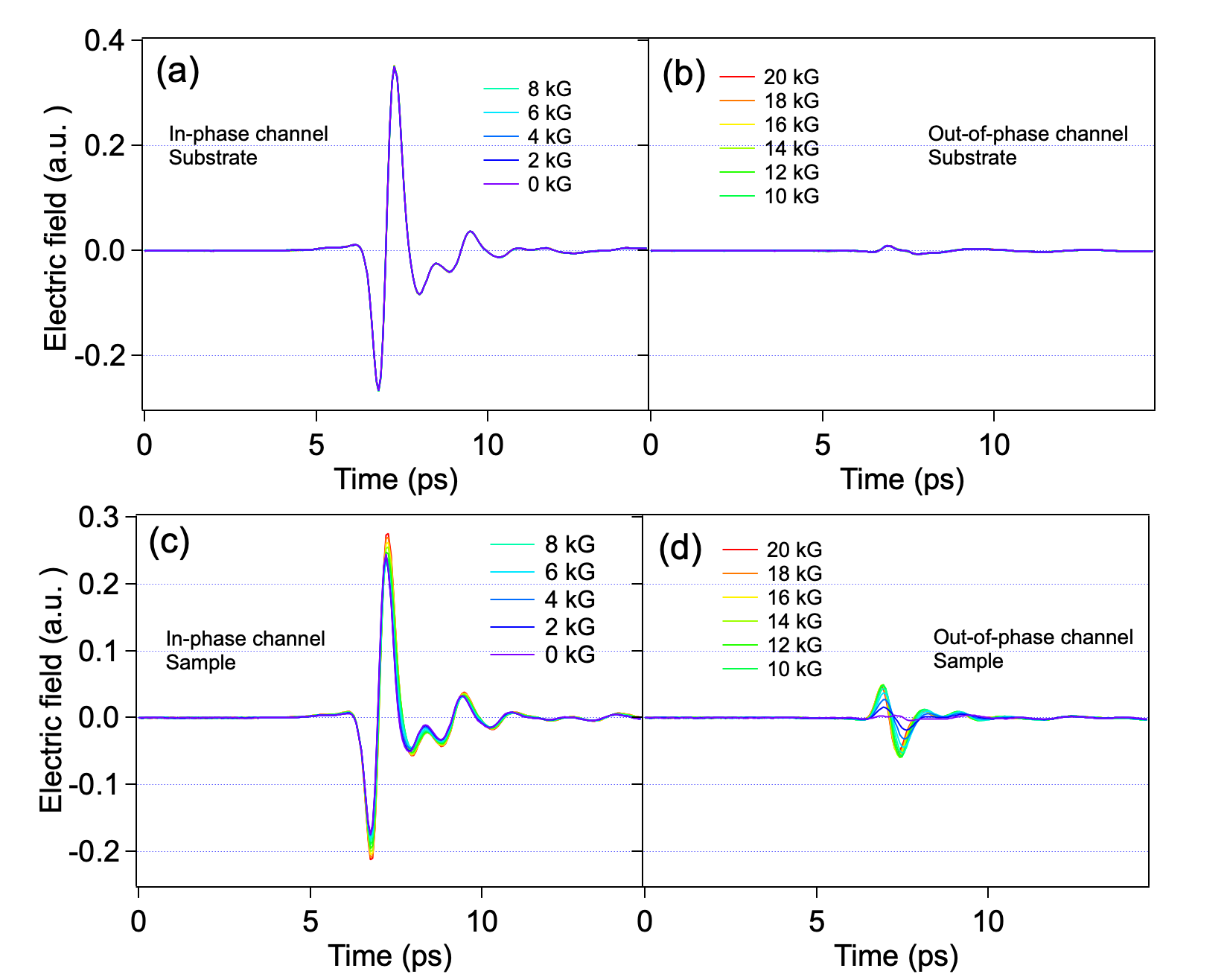}
\caption{(Color online) (a) In-phase and (b) out-of-phase transmitted THz signals of substrate in time domain. (c) In-phase and (d) out-of-phase transmitted THz  signals of Cd$_3$As$_2$ films in time domain. }
\label{trans}
\end{figure*}

\section{S\lowercase{upporting} N\lowercase{ote} 2: R\lowercase{aw time trace data of substrate and sample at} 6 K}

We show the time trace data of the transmitted electric field of the substrate and sample in Figure S1 taken with the fast rotator technique.  In this technique we use a lockin amplifier to lockin in to 2 times the frequency of a spinning polarizer \cite{Morris12a}.  The in-phase signals in Figure S1 give the vertical polarization ($x$) which is also the initial polarization of THz electric field. The out-of-phase signals are the horizontal polarization ($y$). If the signal in this channel is nonzero, it means the polarization of the THz pulse has been rotated by the substrate or sample. In Figure S1a, one can see the in-phase signal of substrate $E_{sub}^{x}$ does not have field dependence. Its out-of-phase signal $E_{sub}^{y}$ (Figure S1b) is negligible and does not have any field evolution. In contrast, one can see the in-phase signal of Cd$_3$As$_2$ $E_{sam}^{x}$ (Figure S1c) has a very clear field dependence. Furthermore, magnetic field introduces notable signal in the out-of-phase channel $E_{sam}^{y}$ (Figure S1d). The field evolution of in-phase and out-of-phase signal of Cd$_3$As$_2$ comes largely from the cyclotron motions of free charges. To perform the analysis in the paper, we Fourier transform this data and ratio the sample transmission to to substrate transmission to get a complex transmission function.  We can use data like that shown in Figure S1 to calculate the transmissions $T_{xx}$ and $T_{xy}$ of Cd$_3$As$_2$ films.  Please see Ref. 1  \cite{Morris12a} for further details on this technique.

\section{S\lowercase{upporting} N\lowercase{ote} 3:  C\lowercase{alculations of optical conductivity in circular polarization basis} }

In the Faraday geometry, the magnetic field is perpendicular to the sample surface and the linear polarization basis is not the eigenbasis of the transmitted THz beam.   If a $C_4$ or $C_3$ symmetry where the rotation axis is along the propogation direction, then the eigenpolarization basis will be circular \cite{armitage2014constraints}.  Then to calculate the optical conductivity, we need to tranform the measured transmissions $T_{xx}$ and $T_{xy}$ to left-hand transmissions $T_r$ and right-hand transmission $T_r$ coefficients through the expression

\begin{equation}
\hat{T}_{cir}={
\left[ \begin{array}{ccc}
T_r  & 0\\
0  & T_l\\

\end{array} 
\right ]}=
{
\left[ \begin{array}{ccc}
T_{xx}+iT_{xy} & 0\\
0 & T_{xx}-iT_{xy}\\

\end{array}
\right ].}
\end{equation}

\noindent We show the magnitude of the the left- and right-hand circularly polarized transmissions at 6 K in Figure S2a and S2b plotted as a function of negative and positive frequencies. One can see the magnetic field introduces notable changes to the transmission in both channels. Especially around 0.7 THz, the phonon mode, indicated by the big dip in the transmission, shows notable shifts in the left hand channel. In contrast, the phonon mode in the right hand channel shows a smaller field dependence. These results are consistent with our analysis of optical conductivity in the main text.
\begin{figure*}[t]
\includegraphics[clip,width=6in]{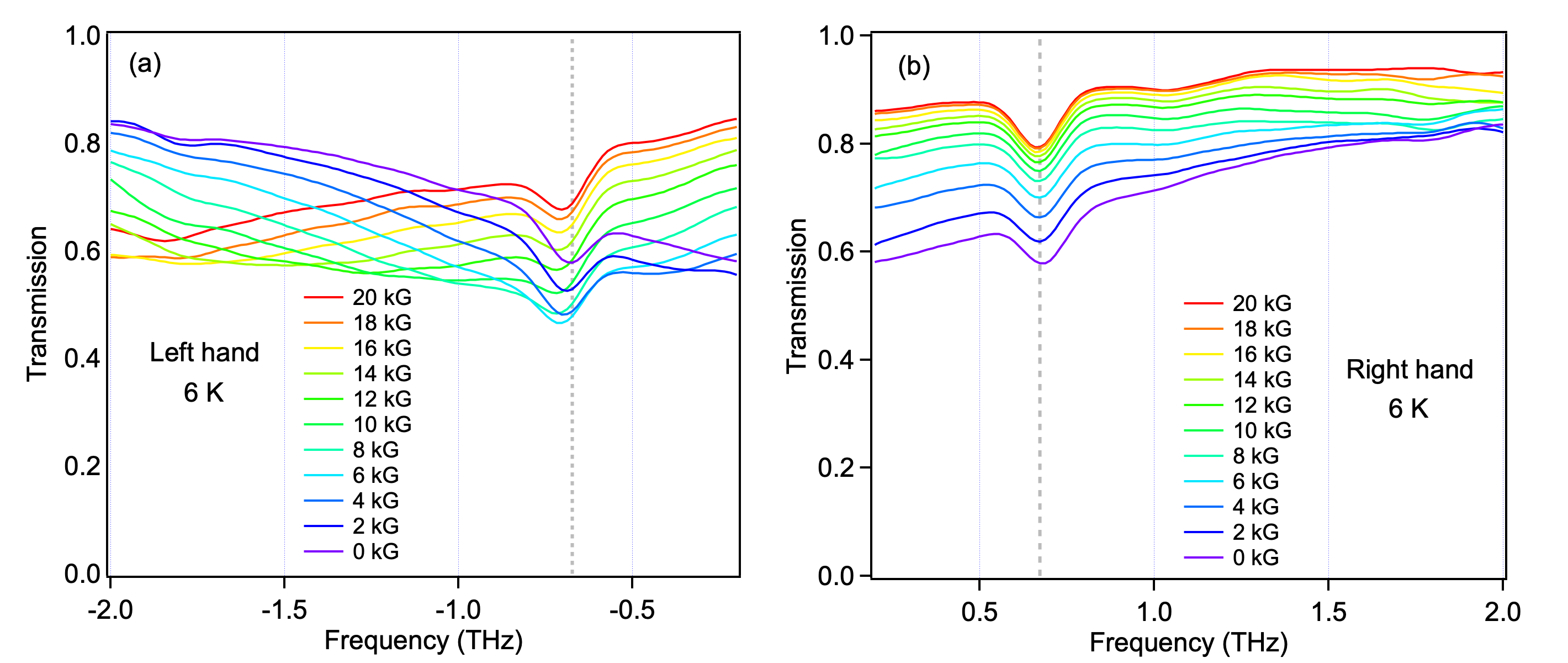}
\caption{(Color online) The magnitude of the the circularly polarized THz transmission of Cd$_3$As$_2$ films in (a) the left-hand channel and in (b) the right-hand channel. The vertical dashed lines represent the position of the optical phonon center frequency at zero field.}
\label{trans}
\end{figure*}

The complex conductivity of the film in circular polarization basis can be directly extracted in the thin-film limit with the expression: $T(\omega) = \frac{1+n}{1+n+Z_{0}d\sigma(\omega)}$exp$[\frac{i\omega}{c}(n-1)\Delta L]$. Here $T(\omega)$ is the left or right hand transmission as referenced to GaAs substrate. $\sigma$($\omega$) is the left or right hand complex optical conductivity. $d$ is the film thickness, and $n$ is the index of refraction of the substrate. $\Delta L$ is the small thickness difference between samples and reference substrates, and $Z_{0}$ is the vacuum impedance, which is approximately 377 $\Omega$.

\section{S\lowercase{upporting} N\lowercase{ote} 4:  F\lowercase{its of optical conductivity} }

To separate electronic and phonon components, we used a Drude/Drude-Lorentz model to fit the complex \textit{r}- and \textit{l}-hand optical conductivities simultaneously. The total THz conductivity is $\sigma_{circ}$ = $\sigma_{Drude}$ + $\sigma_{phonon}$ plus a very weak electron-like oscillator that allows the spectrum to be fit on the positive frequency side at high fields.  Its incorporation has no impact on the conclusions of the paper.  The contribution of the Drude response in magnetic field is the conventional Drude model but including the effect of cyclotron resonance:
\setlength{\abovedisplayskip}{10pt}
\begin{strip}
\begin{equation}
\sigma_{Drude}(\omega)=i\epsilon_0\omega  \Big (\sum_{k=1}^{s}{{-\omega_{pk}^2}\over{-\omega^2-i\omega\Gamma_{pk}+\omega_{cr}\omega}}-(\epsilon_\infty-1) \Big ).
\label{chik}
\end{equation}
\end{strip}

 \noindent In the above expression, $\omega$ runs from positive to negative frequency and $\omega_{cr}$ is the CR frequency.  For hole (electron) carriers, $\omega_{c}$ is positive (negative).  As $\omega_{c}$ goes to zero, this formula automatically recovers the usual Drude form. The expression for the phonon conductivity with the Fano asymmetry is \cite{Phonon_fit_homes_18}:

 \begin{strip}
\begin{equation}
\sigma_{phonon}(\omega)=-i\epsilon_0\omega  \Big [{{\Omega_{p}^2}\over{\omega_0^2-\omega^2 -i\omega\Gamma_{0}}}(1+i\frac{\omega_0}{q\omega})^2+(\frac{\Omega_p}{q\omega})^2 \Big ].
\label{Fano}
\end{equation}
\end{strip}

 \noindent Here, $\Omega_p$ is the phonon's oscillator strength, $\omega_0$ is the phonon's central frequency, $\Gamma_0$ is the phonon linewidth, and $q^{-1}$ is the Fano coupling/asymmetry parameter. As $q^{-1}$ approaches zero, the asymmetry vanishes and the phonon recovers the usual symmetric Lorentzian lineshape. In the main text Figure 2c, we show a Drude-Lorentz fit for the conductivity at 8 kG. One can see that the fit well reproduces the conductivity over the whole frequency region. We also show the pure Drude simulation $\sigma_{Drude}$ without the phonon (green) in the main text Figure 2c. By comparing $\sigma_{cir}$ with $\sigma_{Drude}$, we can see the \textit{L}-hand phonon exhibits a clear Fano shape but the phonon in \textit{R} hand is more symmetric. 

 To plot the optical conductivity of phonon in circular polarization basis, we use the fitting Drude parameters from the whole Drude-Lorentz fit to simulate the Drude response from the pure electronic contribution. Then, we use this pure Drude conductivity $\sigma_{Drude}$ to substract the pure electronic contricutuions from the raw optical conductivity $\sigma_{cir}$

\section{S\lowercase{upporting} N\lowercase{ote} 5: C\lowercase{ircularly polarized eigenstates by applied magnetic field} }

In uniaxial crystals, zone-center phonons polarized in the plane normal to the $z$-axis are doubly degenerate.  Obviously, the detailed mode structure depends on ionic masses, charges and crystal structure, but some aspects can be specified independent of these details.  Here we show quite generically that such phonons become right and left circularly polarized when a magnetic field is applied with a component along the symmetry axis and have energy differences that depend linearly on field  \cite{anastassakis1972morphic}.

In the presence of magnetic field neither the Hamiltonian nor its eigenfunctions have to be real.  Instead the Hamiltonian must satisfy the condition, $ \mathcal{H} (\textbf{B}) =   \mathcal{H}^*(-\textbf{B})) $ where $\text{B}$ is the applied magnetic field.  Therefore the Hamiltonian must have the following form $ \mathcal{H}(\textbf{B})= \mathcal{H}^e(\textbf{B}) + i \mathcal{H}^o(\textbf{B})$ where the first term is real and depends on the magnetic field only to even powers of $\textbf{B}$.  $H^o$ is also real and contains only odd powers of $\textbf{B}$.  Although in principle these terms can contain higher powers of $\textbf{B}$ (and indeed our data shows evidence of this), here we consider only the leading order behavior where only field independent and linear dependencies on $\textbf{B}$ are retained for $\mathcal{H}^e$ and $\mathcal{H}^o$ respectively.

We consider the coupling of magnetic field to optical phonons in a uniaxial crystal structure.   We proceed in usual way for phonons by solving the appropriate classical secular equation for their eigenmodes and \textit{then} quantizing them.   The equation to be solved for a particular doubly degenerate mode $j$ is

\begin{equation}
\mathbf{K}_j \cdot \mathbf{u}_j  =  \omega_j^2     \mathbf{u}_j.
\label{secular}
\end{equation}
Here $\mathbf{K}_j$ is a matrix operator which is an effective spring constant of the mode, $\omega_j$ is the mode frequency, and  $  \mathbf{u}_j $ is the mode amplitude.  The matrix elements of $\mathbf{K}_j$ can be written as
\begin{equation}
K_{\alpha \beta}  =  K_{\alpha \beta}^e(\mathbf{B}) + i  K_{\alpha \beta}^o(\mathbf{B}),
\label{forceconstant}
\end{equation}
where  $K_{\alpha \beta}^e $ and $  K_{\alpha \beta}^o$ contain even and odd powers of $\mathbf{B}$ respectively.    These constants follow from $ \mathcal{H}$ as

\begin{equation}
K_{\alpha \beta}^e =   \Big (  \frac{     \partial^2 \mathcal{H}^e  }{   \partial u_\alpha \partial u_\beta }        \Big  ) ,
\label{derivative1}
\end{equation}

\begin{equation}
K_{\alpha \beta}^o =  \Big(  \frac{     \partial^2 \mathcal{H}^o  }{   \partial u_\alpha \partial u_\beta }        \Big )   \approx  \Big  (  \frac{     \partial^3 \mathcal{H}^o  }{   \partial u_\alpha \partial u_\beta    \partial B_\nu }      \Big ) B_\nu = K_{\alpha \beta \nu} B_\nu,
\label{derivative2}
\end{equation}
with the final approximation following the fact that we retained terms no higher than linear in $\mathbf{B}$ in $\mathcal{H}$.

Onsager reciprocity requires that $K_{\alpha \beta} ( \mathbf{B})  =   K_{ \beta \alpha}    ( - \mathbf{B})   $.  In addition $\mathbf{K}$ must be Hermitian such that the phonon frequencies in the absence of dissipation are real\footnote{Alternatively this follows from requirement of Hermiticity of the Hamiltonian and the definitions in Eqs. \ref{derivative1} and \ref{derivative2}.}.   Together, these give the constraint $  K_{\alpha \beta \nu}   = - K_{\beta \alpha  \nu} $ e.g. that the magnetic-field-induced force constant must be an antisymmetric matrix.   Other contributions based on polar perturbations such as electric field or strain can give a symmetric off-diagonal contribution to the force matrix, but we are not considering these here.   Moreover, the force matrix can be further constrained by spatial symmetry.   Here, the $C_4$ symmetry of Cd$_3$As$_2$ ensures that for fields along $z$ then $K_{xx}^e =  K_{yy}^e$.

In uniaxial crystals like Cd$_3$As$_2$, the only degenerate modes are doubly degenerate and polarized in the plane normal to the $z$-axis.   Therefore we can restrict our analysis to the $x-y$ plane.  One can solve Eq. \ref{secular} above based on Eqs. \ref{forceconstant} - \ref{derivative2} via perturbation theory in the limit of $K_{\alpha \beta}^o  \ll  K_{\alpha \beta}^e$.    One solves

\begin{equation}
\left|\begin{array}{cc} K_{xx}^e  - \omega_j^2 & i K_{xy}^o \\  - i K_{xy}^o & K_{xx}^e  - \omega_j^2 \end{array}\right| = 0.
\label{matrix}
\end{equation}
In this low field limit, one finds that the doubly degenerate lattice normal modes of any uniaxial crystal will split linearly with an applied field e.g.  $\omega_j  \approx \sqrt{K_{xx}^e} \pm  \frac{K_{xyz} B_z}{2 \sqrt{K_{xx}^e}}$.  The eigenstates are $(1/\sqrt{2} ) (1, \pm i ,0) $.    This latter result is generic as applied field removes time-reversal symmetry and the $R$ and $L$ polarization are time reversed versions of each other.  As usual, the quantized vibrations of these classical normal modes are the phonons.  This is a quite general analysis depending only on symmetry and applies to all cases where there is a magnetic field applied along the symmetry direction of a uniaxial crystal.   Therefore under these circumstances we expect that the phonons become circularly polarized and split at lowest order linearly with field and quite generally they can be assigned an orbital phonon magnetic moment.  In our particular case the magnetic field is applied along the 112 direction, and so the field is not purely along the uniaxial direction.   We expect that this introduces a weak $x-y$ anisotropy e.g. magnetic birefringence, an effect that we do not see.

To understand the size of splitting, one must consider a particular mechanism for coupling the magnetic field to the ions.  In the simplest case of a non-magnetic insulator, the only force on the ions is the Lorentz force.   Dynamical matrix calculations can be done that depend upon lattice symmetry, and ionic masses and charges as input parameters \cite{anastassakis1972morphic,vineyard1985effect}, but in general the scale of  $ \sqrt{K_{xy}^o}$ is of order the ion cyclotron frequency times half the zero field phonon frequency $\omega_0$. Therefore $\omega_j  \approx \omega_0 \pm \alpha eB/M$, where $M$ is an effective ionic mass and $\alpha$ is a number of order unity.   For Cd$_3$As$_2$ this gives a splitting between the two branches of order 3 MHz at 1 Tesla, the small size of which is consistent with ab initio calculations \cite{juraschek2018orbital}, but is obviously much smaller than we observe.  Paramagnets show an effect with precisely the same symmetry as discussed here, but which is dramatically enhanced over the simple Lorentz force on ions mechanism \cite{schaack1976observation}.

\begin{figure*}[t]
\includegraphics[clip,width=3in]{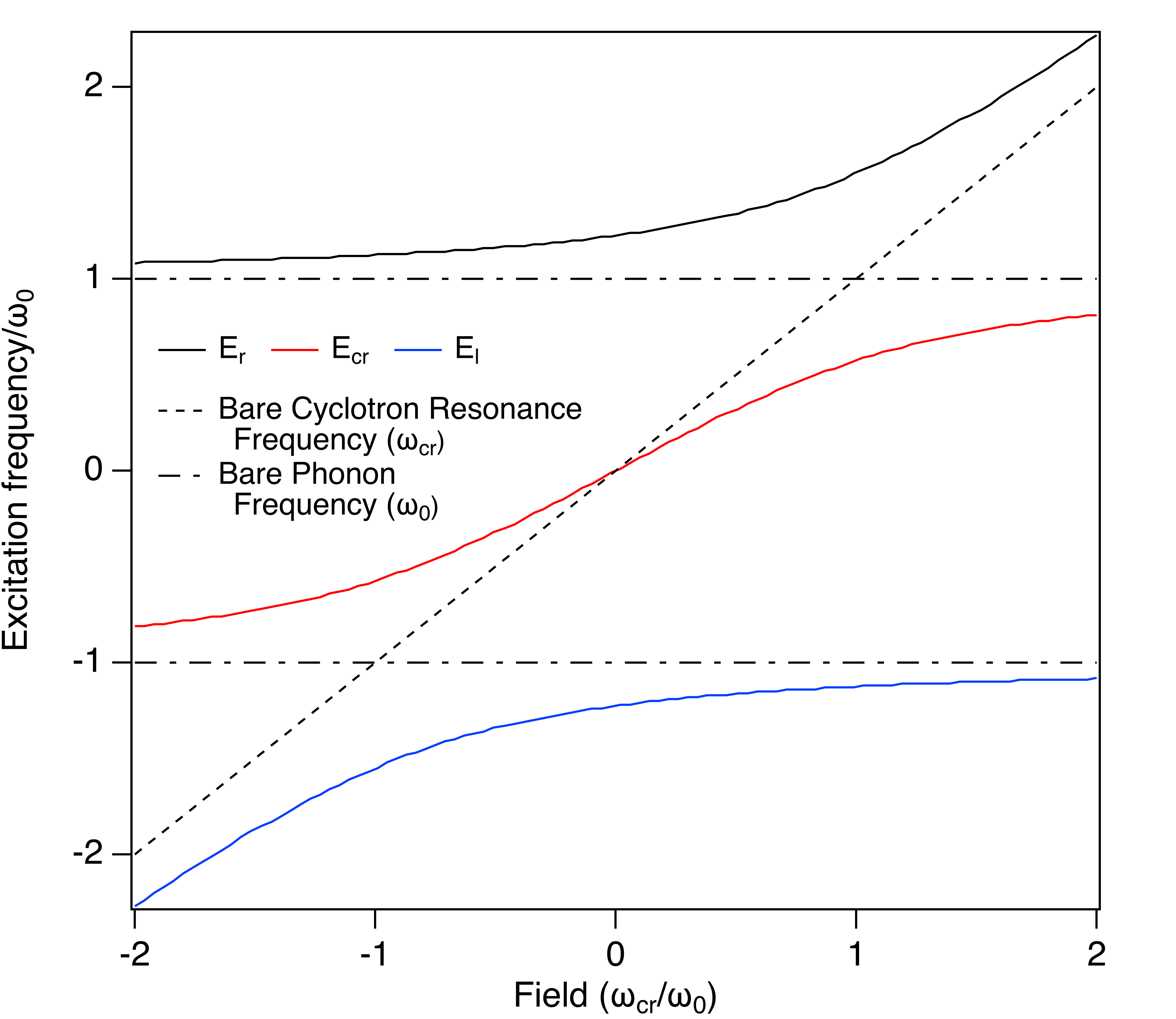}
\includegraphics[clip,width=3in]{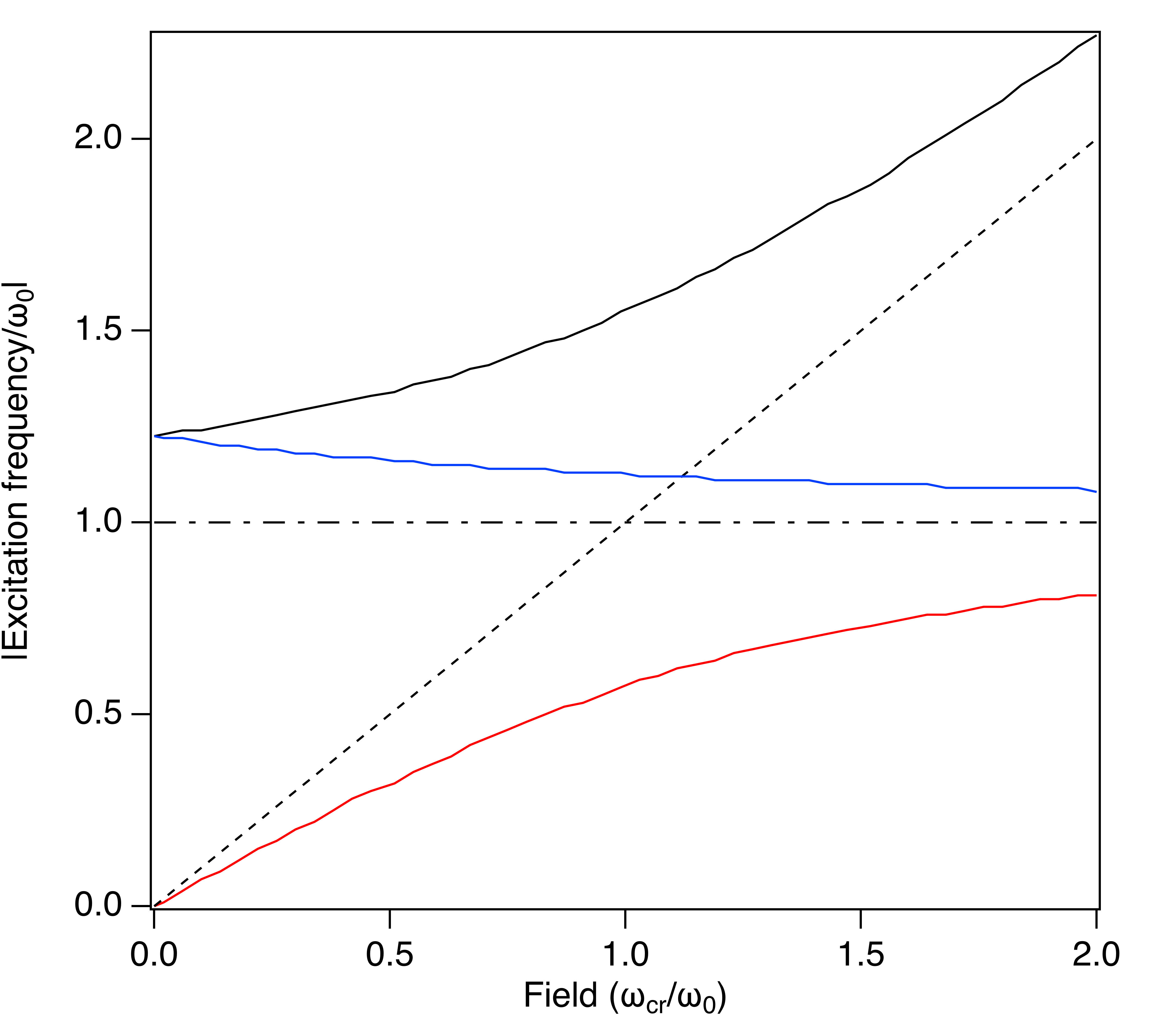}
\caption{Eigenvalues of Hamiltonian Eq. \ref{matrix} in units of $ \omega_{0}$ as a function of magnetic field in units of $\omega_{cr} / \omega_{0}$.   Here we have chosen $g_{ep} = 0.5 \hbar \omega_{0}$.   On the left is plotted eigenfrequencies as positive and negative frequencies that express r and l polarized modes. The dashed diagonal line is the bare cyclotron frequency.   The same data is plotted on the right with the absolute value of frequency.   This highlights the fact that the phonon eigenfrequencies split linearly with applied field. }
\label{Cyclo}
\end{figure*}

In the present case of a non-magnetic semimetallic system  we can consider a model where we couple a cyclotron mode to $r$ and $l$ polarized optical phonons though a phenomenological electron-phonon coupling.  This is a stripped down version of the models used in Ref. \cite{xiaoguang1985two,peeters1986cyclotron} for polar semiconductors and in Ref. \cite{goerbig2011electronic,goerbig2007filling} for graphene.   It can model the dependence of the eigenmodes as a function of frequency.  One can consider a Hamiltonian that can be written as

\begin{equation}
H = \left[\begin{array}{ccc} \hbar \omega_{cr} & g_{ep} & g_{ep} \\  g_{ep} & \hbar \omega_{0} & 0 \\ g_{ep} &  0 & - \hbar \omega_{0} \end{array}\right],
\label{matrix}
\end{equation}
in which $\omega_{cr}$ is the bare cyclotron frequency of the charge carriers $eB/m$, $\pm \omega_{0}$ are the r and l optical Einstein phonons expressed as positive and negative frequencies, and $g_{ep}$ is the electron-phonon coupling constant.  Unfortunately, even for this simple Hamiltonian the energy eigenvalue structure is complicated and giving the explicit eigenvalues and eigenvectors for the full range of Hamiltonian parameters would not be very illuminating. We show them graphically in Figure \ref{Cyclo} as a function of $\omega_{cr} / \omega_{0}$ for the case of $g_{ep} = 0.5 \hbar \omega_{0}$.   One can see that at zero field the phonon-like modes have their energies renormalized by coupling to the cyclotron mode.   Their energies progressively change as the field is tuned and the ``bare" cyclotron resonance frequency is tuned through the phonon energies.   Moreover, the cyclotron resonance is changed via the coupling to the phonons to have a weaker slope in the $\omega_{cr} \rightarrow 0$ limit as compared to the bare cyclotron resonance.

To gain further insight we diagonalize the Hamiltonian in the limit of $\omega_{cr} = 0$.   In this limit one finds eigenvalues of $E_{cr} = 0$ and $E_{r,l} = \pm \sqrt{ (\hbar \omega_{0})^2 + g_{ep}^2}$.   Then we evaluate the dependence at finite field including $\hbar \omega_{cr} $ as a perturbation.   The three eigenvectors (unnormalized) in this limit are

\begin{strip}
\begin{align}
| cr' \rangle &= \Big[ \frac{\hbar \omega_{0}}{ g_{ep}}, -1, 1\Big]  \\ 
|r' \rangle &= \Big[ \frac{\hbar \omega_{0} - \sqrt{2 g_{ep}^2 + (\hbar \omega_{0})^2 } }{ g_{ep}}, 
  \frac{g_{ep}^2 + (\hbar \omega_{0})^2 - \hbar \omega_{0}  \sqrt{2 g_{ep}^2 +  (\hbar \omega_{0})^2  } }{g_{ep}^2}, 1 \Big] \\
| l'\rangle &= \Big[ \frac{\hbar \omega_{0} + \sqrt{2 g_{ep}^2 + (\hbar \omega_{0})^2 } }{ g_{ep}}, 
  \frac{g_{ep}^2 + (\hbar \omega_{0})^2 + \hbar \omega_{0}  \sqrt{2 g_{ep}^2 +  (\hbar \omega_{0})^2  } }{g_{ep}^2}, 1 \Big].
 \end{align}
 \end{strip}
 
 Now we can perturb and calculate the first order energy shifts with the finite field term, which is 
 
 \begin{equation}
H_{cr} = \left[\begin{array}{ccc} \hbar \omega_{cr} & 0 & 0 \\  0 & 0 & 0 \\ 0 &  0 & 0 \end{array}\right].
\label{matrix}
\end{equation}

The energy shifts under applied field are 

\begin{align}
\Delta E_{cr} &= \hbar \omega_{cr} \frac{ 1 }{1 + 2 g_{ep}^2/ (\hbar \omega_{0})^2  } , \\
\Delta E_{r,l} &= \hbar \omega_{cr} \frac{g_{ep}^2}{(\hbar \omega_{0})^2  + g_{ep}^2 }.
\end{align}

With the assignment of $\lambda = 2 g_{ep}^2 / (\hbar \omega_{0})^2 $ as the dimensionless electron-phonon coupling constant one finds  

\begin{align}
E_{cr} = \hbar \omega_{cr} \frac{ 1 }{1 + \lambda} , \\
E_{r,l} = \pm \hbar \omega_{0} \sqrt{ 1 + \lambda/2} + \hbar \omega_{cr} \frac{ \lambda/2 }{ 1 + \lambda }.
\end{align}

Eq. 16 is the conventional result for the renormalization of the cyclotron resonance frequency due to electron-phonon coupling with a dimensionless electron-phonon coupling constant $\lambda$ \cite{xiaoguang1985two,peeters1986cyclotron,goerbig2011electronic,goerbig2007filling}.   Eq. 17 shows that size of the phonon magnetic moment is directly set by the electron phonon coupling.  This gives the remarkable result that the energy difference between the phonon splittings divided by the cyclotron resonance frequency is precisely the electron-phonon coupling constant $\lambda$.   This minimal model gives a novel mechanism for generation of large phonon magnetic moments through electron-phonon coupling.  As discussed in the main text, as applied to our data this analysis gives a dimensionless electron-phonon coupling constant of approximately 0.09.

Away from zero field, this minimal model is in reasonable correspondence with our experimental results, although there are some differences as well.  In particular the linear and equal splitting of the phonons near zero field, and also the stronger dependence of the upper phonon branch at larger fields is captured.   As mentioned, it also correctly captures the renormalization of the electron mass that appears in the cyclotron mode. It does NOT properly describe the eventual dependence of the upper phonon branch that in our model smoothly evolves back into a cyclotron resonance-like mode.   In our experimental data, the cyclotron resonance mode increases essentially unabated through the phonon spectral region.   This may be due to relative spectral width of the cyclotron resonance peak as compared to the phonon e.g. only a small spectral slice of the cyclotron resonance is resonant with the phonon at any particular field.   Moreover the detailed evolution of the modes is undoubtedly more complicated than in our minimal model, as there are many more phonons (particularly Raman active ones) in the energy region of the prominent IR mode.   These will serve to "pin" the field evolution of the phonon modes in certain bands as observed in InSb \cite{mccombe1968effects}.

\bibliography{Quadratic}

\providecommand{\latin}[1]{#1}
\providecommand*\mcitethebibliography{\thebibliography}
\csname @ifundefined\endcsname{endmcitethebibliography}
  {\let\endmcitethebibliography\endthebibliography}{}
\begin{mcitethebibliography}{33}
\providecommand*\natexlab[1]{#1}
\providecommand*\mciteSetBstSublistMode[1]{}
\providecommand*\mciteSetBstMaxWidthForm[2]{}
\providecommand*\mciteBstWouldAddEndPuncttrue
  {\def\EndOfBibitem{\unskip.}}
\providecommand*\mciteBstWouldAddEndPunctfalse
  {\let\EndOfBibitem\relax}
\providecommand*\mciteSetBstMidEndSepPunct[3]{}
\providecommand*\mciteSetBstSublistLabelBeginEnd[3]{}
\providecommand*\EndOfBibitem{}
\mciteSetBstSublistMode{f}
\mciteSetBstMaxWidthForm{subitem}{(\alph{mcitesubitemcount})}
\mciteSetBstSublistLabelBeginEnd
  {\mcitemaxwidthsubitemform\space}
  {\relax}
  {\relax}

\bibitem[Armitage \latin{et~al.}(2018)Armitage, Mele, and Vishwanath]{NPA18}
Armitage,~N.~P.; Mele,~E.~J.; Vishwanath,~A. Weyl and \text{Dirac} semimetals
  in three-dimensional solids. \emph{Rev. Mod. Phys.} \textbf{2018}, \emph{90},
  015001\relax
\mciteBstWouldAddEndPuncttrue
\mciteSetBstMidEndSepPunct{\mcitedefaultmidpunct}
{\mcitedefaultendpunct}{\mcitedefaultseppunct}\relax
\EndOfBibitem
\bibitem[Song \latin{et~al.}(2016)Song, Zhao, Fang, and
  Dai]{Phonon_chiral_2016}
Song,~Z.; Zhao,~J.; Fang,~Z.; Dai,~X. Detecting the chiral magnetic effect by
  lattice dynamics in \text{Weyl} semimetals. \emph{Phys. Rev. B}
  \textbf{2016}, \emph{94}, 214306\relax
\mciteBstWouldAddEndPuncttrue
\mciteSetBstMidEndSepPunct{\mcitedefaultmidpunct}
{\mcitedefaultendpunct}{\mcitedefaultseppunct}\relax
\EndOfBibitem
\bibitem[Rinkel \latin{et~al.}(2017)Rinkel, Lopes, and
  Garate]{Phonon_chiral_2017}
Rinkel,~P.; Lopes,~P. L.~S.; Garate,~I. Signatures of the Chiral Anomaly in
  Phonon Dynamics. \emph{Phys. Rev. Lett.} \textbf{2017}, \emph{119},
  107401\relax
\mciteBstWouldAddEndPuncttrue
\mciteSetBstMidEndSepPunct{\mcitedefaultmidpunct}
{\mcitedefaultendpunct}{\mcitedefaultseppunct}\relax
\EndOfBibitem
\bibitem[Liu and Shi(2017)Liu, and Shi]{Circular_phonon_Dichroism_2017}
Liu,~D.; Shi,~J. Circular Phonon Dichroism in \text{Weyl} Semimetals.
  \emph{Phys. Rev. Lett.} \textbf{2017}, \emph{119}, 075301\relax
\mciteBstWouldAddEndPuncttrue
\mciteSetBstMidEndSepPunct{\mcitedefaultmidpunct}
{\mcitedefaultendpunct}{\mcitedefaultseppunct}\relax
\EndOfBibitem
\bibitem[Kuroda and et~al.(2015)Kuroda, and et~al.]{Mn3Sn_photoemission_2017}
Kuroda,~K.; et~al., Evidence for magnetic \text{Weyl} fermions in a correlated
  metal. \emph{Nat. Mater.} \textbf{2015}, \emph{16}, 1090\relax
\mciteBstWouldAddEndPuncttrue
\mciteSetBstMidEndSepPunct{\mcitedefaultmidpunct}
{\mcitedefaultendpunct}{\mcitedefaultseppunct}\relax
\EndOfBibitem
\bibitem[Barnes \latin{et~al.}(1991)Barnes, Nicholas, Peeters, Wu, Devreese,
  Singleton, Langerak, Harris, and Foxon]{Magneto_phonon_1991}
Barnes,~D.~J.; Nicholas,~R.~J.; Peeters,~F.~M.; Wu,~X.-G.; Devreese,~J.~T.;
  Singleton,~J.; Langerak,~C. J. G.~M.; Harris,~J.~J.; Foxon,~C.~T. Observation
  of optically detected magnetophonon resonance. \emph{Phys. Rev. Lett.}
  \textbf{1991}, \emph{66}, 794--797\relax
\mciteBstWouldAddEndPuncttrue
\mciteSetBstMidEndSepPunct{\mcitedefaultmidpunct}
{\mcitedefaultendpunct}{\mcitedefaultseppunct}\relax
\EndOfBibitem
\bibitem[Wang \latin{et~al.}(1997)Wang, Nickel, McCombe, Peeters, Shi, Hai, Wu,
  Eustis, and Schaff]{GaAs_quantum_well_phonon_1997}
Wang,~Y.~J.; Nickel,~H.~A.; McCombe,~B.~D.; Peeters,~F.~M.; Shi,~J.~M.;
  Hai,~G.~Q.; Wu,~X.-G.; Eustis,~T.~J.; Schaff,~W. Resonant Magnetopolaron
  Effects due to Interface Phonons in \text{GaAs}/\text{AlGaAs} Multiple
  Quantum Well Structures. \emph{Phys. Rev. Lett.} \textbf{1997}, \emph{79},
  3226--3229\relax
\mciteBstWouldAddEndPuncttrue
\mciteSetBstMidEndSepPunct{\mcitedefaultmidpunct}
{\mcitedefaultendpunct}{\mcitedefaultseppunct}\relax
\EndOfBibitem
\bibitem[Juraschek and Spaldin(2019)Juraschek, and
  Spaldin]{juraschek2018orbital}
Juraschek,~D.~M.; Spaldin,~N.~A. Orbital magnetic moments of phonons.
  \emph{Physical Review Materials} \textbf{2019}, \emph{3}, 064405\relax
\mciteBstWouldAddEndPuncttrue
\mciteSetBstMidEndSepPunct{\mcitedefaultmidpunct}
{\mcitedefaultendpunct}{\mcitedefaultseppunct}\relax
\EndOfBibitem
\bibitem[Ali \latin{et~al.}(2014)Ali, Gibson, Jeon, Zhou, Yazdani, and
  Cava]{ali2014crystal}
Ali,~M.~N.; Gibson,~Q.; Jeon,~S.; Zhou,~B.~B.; Yazdani,~A.; Cava,~R.~J. The
  crystal and electronic structures of \text{Cd$_3$As$_2$}, the
  three-dimensional electronic analogue of graphene. \emph{Inorganic chemistry}
  \textbf{2014}, \emph{53}, 4062--4067\relax
\mciteBstWouldAddEndPuncttrue
\mciteSetBstMidEndSepPunct{\mcitedefaultmidpunct}
{\mcitedefaultendpunct}{\mcitedefaultseppunct}\relax
\EndOfBibitem
\bibitem[Schumann \latin{et~al.}(2016)Schumann, Goyal, Kim, and
  Stemmer]{Timo_CdAs_growth_16}
Schumann,~T.; Goyal,~M.; Kim,~H.; Stemmer,~S. Molecular beam epitaxy of
  \text{Cd$_3$As$_2$} on a \text{III-V} substrate. \emph{APL Mater.}
  \textbf{2016}, \emph{4}, 126110\relax
\mciteBstWouldAddEndPuncttrue
\mciteSetBstMidEndSepPunct{\mcitedefaultmidpunct}
{\mcitedefaultendpunct}{\mcitedefaultseppunct}\relax
\EndOfBibitem
\bibitem[Nakazawa \latin{et~al.}(2018)Nakazawa, Uchida, Nishihaya, Kriener,
  Kozuka, Taguchi, and Kawasaki]{CdAs_MBE_2}
Nakazawa,~Y.; Uchida,~M.; Nishihaya,~S.; Kriener,~M.; Kozuka,~Y.; Taguchi,~Y.;
  Kawasaki,~M. Structural characterisation of high-mobility \text{Cd$_3$As$_2$}
  films crystallised on \text{SrTiO$_3$}. \emph{Sci. Rep} \textbf{2018},
  \emph{8}, 2244\relax
\mciteBstWouldAddEndPuncttrue
\mciteSetBstMidEndSepPunct{\mcitedefaultmidpunct}
{\mcitedefaultendpunct}{\mcitedefaultseppunct}\relax
\EndOfBibitem
\bibitem[Cheng \latin{et~al.}(2019)Cheng, Taylor, Folkes, Rong, and
  Armitage]{TCI_Bing_PRL}
Cheng,~B.; Taylor,~P.; Folkes,~P.; Rong,~C.; Armitage,~N.~P. Magnetoterahertz
  Response and \text{Faraday} Rotation from Massive \text{Dirac} Fermions in
  the Topological Crystalline Insulator \text{Pb$_{0.5}$Sn$_{0.5}$Te}.
  \emph{Phys. Rev. Lett.} \textbf{2019}, \emph{122}, 097401\relax
\mciteBstWouldAddEndPuncttrue
\mciteSetBstMidEndSepPunct{\mcitedefaultmidpunct}
{\mcitedefaultendpunct}{\mcitedefaultseppunct}\relax
\EndOfBibitem
\bibitem[Morris \latin{et~al.}(2012)Morris, Aguilar, Stier, and
  Armitage]{Morris12a}
Morris,~C.~M.; Aguilar,~R.~V.; Stier,~A.~V.; Armitage,~N.~P. Polarization
  modulation time-domain terahertz polarimetry. \emph{Opt. Express}
  \textbf{2012}, \emph{20}, 12303\relax
\mciteBstWouldAddEndPuncttrue
\mciteSetBstMidEndSepPunct{\mcitedefaultmidpunct}
{\mcitedefaultendpunct}{\mcitedefaultseppunct}\relax
\EndOfBibitem
\bibitem[Goyal \latin{et~al.}(2018)Goyal, Galletti, Salmani-Rezaie, Schumann,
  Kealhofer, and Stemmer]{goyal2018thickness}
Goyal,~M.; Galletti,~L.; Salmani-Rezaie,~S.; Schumann,~T.; Kealhofer,~D.~A.;
  Stemmer,~S. Thickness dependence of the quantum \text{Hall} effect in films
  of the three-dimensional \text{Dirac} semimetal \text{Cd$_3$As$_2$}.
  \emph{APL Materials} \textbf{2018}, \emph{6}, 026105\relax
\mciteBstWouldAddEndPuncttrue
\mciteSetBstMidEndSepPunct{\mcitedefaultmidpunct}
{\mcitedefaultendpunct}{\mcitedefaultseppunct}\relax
\EndOfBibitem
\bibitem[Schumann \latin{et~al.}(2017)Schumann, Goyal, Kealhofer, and
  Stemmer]{Dirac_Timo_prb}
Schumann,~T.; Goyal,~M.; Kealhofer,~D.~A.; Stemmer,~S. Negative
  magnetoresistance due to conductivity fluctuations in films of the
  topological semimetal \text{Cd$_3$As$_2$}. \emph{Phys. Rev. B} \textbf{2017},
  \emph{95}, 241113\relax
\mciteBstWouldAddEndPuncttrue
\mciteSetBstMidEndSepPunct{\mcitedefaultmidpunct}
{\mcitedefaultendpunct}{\mcitedefaultseppunct}\relax
\EndOfBibitem
\bibitem[Liang \latin{et~al.}(2015)Liang, Gibson, Ali, Liu, Cava, and
  Ong]{Cd3As2_magneto_resis_2014}
Liang,~T.; Gibson,~G.; Ali,~M.; Liu,~M.; Cava,~R.~J.; Ong,~N.~P. Ultrahigh
  mobility and giant magnetoresistance in the \text{Dirac} semimetal
  \text{Cd3As2}. \emph{Nat. Mater.} \textbf{2015}, \emph{14}, 14933\relax
\mciteBstWouldAddEndPuncttrue
\mciteSetBstMidEndSepPunct{\mcitedefaultmidpunct}
{\mcitedefaultendpunct}{\mcitedefaultseppunct}\relax
\EndOfBibitem
\bibitem[Leahy \latin{et~al.}(2018)Leahy, Lin, Siegfried, Treglia, Song,
  Nandkishore, and Lee]{magnetoresistivity_pnas}
Leahy,~I.~A.; Lin,~Y.-P.; Siegfried,~P.~E.; Treglia,~A.~C.; Song,~J. C.~W.;
  Nandkishore,~R.~M.; Lee,~M. Nonsaturating large magnetoresistance in
  semimetals. \emph{Proceedings of the National Academy of Sciences}
  \textbf{2018}, \emph{115}, 10570--10575\relax
\mciteBstWouldAddEndPuncttrue
\mciteSetBstMidEndSepPunct{\mcitedefaultmidpunct}
{\mcitedefaultendpunct}{\mcitedefaultseppunct}\relax
\EndOfBibitem
\bibitem[Tang \latin{et~al.}(2009)Tang, Zhang, Park, Geng, Girit, Hao, Martin,
  Zettl, Crommie, Louie, Shen, and Wang]{Bilayer-graphene_Fano_2009}
Tang,~T.-T.; Zhang,~Y.; Park,~C.-H.; Geng,~B.; Girit,~C.; Hao,~Z.;
  Martin,~M.~C.; Zettl,~A.; Crommie,~M.~F.; Louie,~S.~G.; Shen,~Y.~R.; Wang,~F.
  A tunable phonon–exciton \text{Fano} system in bilayer graphene. \emph{Nat.
  Nanotech.} \textbf{2009}, \emph{5}, 32\relax
\mciteBstWouldAddEndPuncttrue
\mciteSetBstMidEndSepPunct{\mcitedefaultmidpunct}
{\mcitedefaultendpunct}{\mcitedefaultseppunct}\relax
\EndOfBibitem
\bibitem[Schaack(1976)]{schaack1976observation}
Schaack,~G. Observation of circularly polarized phonon states in an external
  magnetic field. \emph{Journal of Physics C: Solid State Physics}
  \textbf{1976}, \emph{9}, L297\relax
\mciteBstWouldAddEndPuncttrue
\mciteSetBstMidEndSepPunct{\mcitedefaultmidpunct}
{\mcitedefaultendpunct}{\mcitedefaultseppunct}\relax
\EndOfBibitem
\bibitem[Anastassakis \latin{et~al.}(1972)Anastassakis, Burstein, Maradudin,
  and Minnick]{anastassakis1972morphic}
Anastassakis,~E.; Burstein,~E.; Maradudin,~A.; Minnick,~R. Morphic
  effects—\text{III}. Effects of an external magnetic field on the long
  wavelength optical phonons. \emph{Journal of Physics and Chemistry of Solids}
  \textbf{1972}, \emph{33}, 519--531\relax
\mciteBstWouldAddEndPuncttrue
\mciteSetBstMidEndSepPunct{\mcitedefaultmidpunct}
{\mcitedefaultendpunct}{\mcitedefaultseppunct}\relax
\EndOfBibitem
\bibitem[Qin \latin{et~al.}(2012)Qin, Zhou, and Shi]{phonon_Berry_2011}
Qin,~T.; Zhou,~J.; Shi,~J. Berry curvature and the phonon \text{Hall} effect.
  \emph{Phys. Rev. B} \textbf{2012}, \emph{86}, 104305\relax
\mciteBstWouldAddEndPuncttrue
\mciteSetBstMidEndSepPunct{\mcitedefaultmidpunct}
{\mcitedefaultendpunct}{\mcitedefaultseppunct}\relax
\EndOfBibitem
\bibitem[Zhang and Niu(2015)Zhang, and Niu]{zhang2015chiral}
Zhang,~L.; Niu,~Q. Chiral Phonons at High-Symmetry Points in Monolayer
  Hexagonal Lattices. \emph{Phys. Rev. Lett.} \textbf{2015}, \emph{115},
  115502\relax
\mciteBstWouldAddEndPuncttrue
\mciteSetBstMidEndSepPunct{\mcitedefaultmidpunct}
{\mcitedefaultendpunct}{\mcitedefaultseppunct}\relax
\EndOfBibitem
\bibitem[Zhu \latin{et~al.}(2018)Zhu, Yi, Li, Xiao, Zhang, Yang, Kaindl, Li,
  Wang, and Zhang]{zhu2018observation}
Zhu,~H.; Yi,~J.; Li,~M.-Y.; Xiao,~J.; Zhang,~L.; Yang,~C.-W.; Kaindl,~R.~A.;
  Li,~L.-J.; Wang,~Y.; Zhang,~X. Observation of chiral phonons. \emph{Science}
  \textbf{2018}, \emph{359}, 579--582\relax
\mciteBstWouldAddEndPuncttrue
\mciteSetBstMidEndSepPunct{\mcitedefaultmidpunct}
{\mcitedefaultendpunct}{\mcitedefaultseppunct}\relax
\EndOfBibitem
\bibitem[Xiaoguang \latin{et~al.}(1985)Xiaoguang, Peeters, and
  Devreese]{xiaoguang1985two}
Xiaoguang,~W.; Peeters,~F.; Devreese,~J. Two-dimensional polaron in a magnetic
  field. \emph{Physical Review B} \textbf{1985}, \emph{32}, 7964\relax
\mciteBstWouldAddEndPuncttrue
\mciteSetBstMidEndSepPunct{\mcitedefaultmidpunct}
{\mcitedefaultendpunct}{\mcitedefaultseppunct}\relax
\EndOfBibitem
\bibitem[Peeters \latin{et~al.}(1986)Peeters, Xiaoguang, and
  Devreese]{peeters1986cyclotron}
Peeters,~F.; Xiaoguang,~W.; Devreese,~J. Cyclotron mass of a polaron in two
  dimensions. \emph{Physical Review B} \textbf{1986}, \emph{34}, 1160\relax
\mciteBstWouldAddEndPuncttrue
\mciteSetBstMidEndSepPunct{\mcitedefaultmidpunct}
{\mcitedefaultendpunct}{\mcitedefaultseppunct}\relax
\EndOfBibitem
\bibitem[Goerbig(2011)]{goerbig2011electronic}
Goerbig,~M. Electronic properties of graphene in a strong magnetic field.
  \emph{Reviews of Modern Physics} \textbf{2011}, \emph{83}, 1193\relax
\mciteBstWouldAddEndPuncttrue
\mciteSetBstMidEndSepPunct{\mcitedefaultmidpunct}
{\mcitedefaultendpunct}{\mcitedefaultseppunct}\relax
\EndOfBibitem
\bibitem[Goerbig \latin{et~al.}(2007)Goerbig, Fuchs, Kechedzhi, and
  Fal’ko]{goerbig2007filling}
Goerbig,~M.; Fuchs,~J.-N.; Kechedzhi,~K.; Fal’ko,~V.~I.
  Filling-factor-dependent magnetophonon resonance in graphene. \emph{Physical
  Review Letters} \textbf{2007}, \emph{99}, 087402\relax
\mciteBstWouldAddEndPuncttrue
\mciteSetBstMidEndSepPunct{\mcitedefaultmidpunct}
{\mcitedefaultendpunct}{\mcitedefaultseppunct}\relax
\EndOfBibitem
\bibitem[Hubener \latin{et~al.}(2017)Hubener, Sentef, Giovannini, Kemper, and
  Rubio]{light_induced_topo_phase_2017}
Hubener,~H.; Sentef,~M.~A.; Giovannini,~U.~D.; Kemper,~A.~F.; Rubio,~A.
  Creating stable Floquet-\text{Weyl} semimetals by laser-driving of 3\text{D}
  \text{Dirac} materials. \emph{Nat. Commun.} \textbf{2017}, \emph{8},
  13940\relax
\mciteBstWouldAddEndPuncttrue
\mciteSetBstMidEndSepPunct{\mcitedefaultmidpunct}
{\mcitedefaultendpunct}{\mcitedefaultseppunct}\relax
\EndOfBibitem
\bibitem[Armitage(2014)]{armitage2014constraints}
Armitage,~N.~P. Constraints on Jones transmission matrices from time-reversal
  invariance and discrete spatial symmetries. \emph{Phys. Rev. B}
  \textbf{2014}, \emph{90}, 035135\relax
\mciteBstWouldAddEndPuncttrue
\mciteSetBstMidEndSepPunct{\mcitedefaultmidpunct}
{\mcitedefaultendpunct}{\mcitedefaultseppunct}\relax
\EndOfBibitem
\bibitem[Homes \latin{et~al.}(2018)Homes, Dai, Akrap, Bud'ko, and
  Canfield]{Phonon_fit_homes_18}
Homes,~C.~C.; Dai,~Y.~M.; Akrap,~A.; Bud'ko,~S.~L.; Canfield,~P.~C. Vibrational
  anomalies in \text{BaFe$_2$As$_2$} (\text{A=Ca, Sr, and Ba}) single crystals.
  \emph{Phys. Rev. B} \textbf{2018}, \emph{98}, 035103\relax
\mciteBstWouldAddEndPuncttrue
\mciteSetBstMidEndSepPunct{\mcitedefaultmidpunct}
{\mcitedefaultendpunct}{\mcitedefaultseppunct}\relax
\EndOfBibitem
\bibitem[Vineyard(1985)]{vineyard1985effect}
Vineyard,~G.~H. Effect of a magnetic field on the vibrations of an ionic
  lattice. \emph{Physical Review B} \textbf{1985}, \emph{31}, 814\relax
\mciteBstWouldAddEndPuncttrue
\mciteSetBstMidEndSepPunct{\mcitedefaultmidpunct}
{\mcitedefaultendpunct}{\mcitedefaultseppunct}\relax
\EndOfBibitem
\bibitem[McCombe and Kaplan(1968)McCombe, and Kaplan]{mccombe1968effects}
McCombe,~B.; Kaplan,~R. Effects of electron-optical-phonon interaction in the
  combined resonance spectra of insb. \emph{Physical Review Letters}
  \textbf{1968}, \emph{21}, 756\relax
\mciteBstWouldAddEndPuncttrue
\mciteSetBstMidEndSepPunct{\mcitedefaultmidpunct}
{\mcitedefaultendpunct}{\mcitedefaultseppunct}\relax
\EndOfBibitem
\end{mcitethebibliography}

\end{document}